# Observation of Giant Exchange Bias and Topological Hall Effect in Manganese Nitride Films

Meng Meng[1,2], Shuwei Li[1], Mohammad Saghayezhian[2], E. W. Plummer[2], Rongying Jin[2],

1. *State Key Laboratory of Optoelectronic Materials and Technologies, School of Materials Science & Engineering, Sun Yat-sen University, Guangzhou 510275, China*
2. *Department of Physics & Astronomy, Louisiana State University, Baton Rouge, Louisiana 70803, USA*

**Abstract:** Magnetic and magneto-transport properties of manganese nitride films grown by molecular beam epitaxy have been investigated. Due to the mixed ferrimagnetic (FI) phase (ε-phase with $T_{FI}$ ~ 738 K) and the antiferromagnetic phase (ζ-phase with $T_N$ ~ 273 K), we observe magnetization hysteresis loops with non-zero exchange bias below $T_N$, reaching ~ 0.22 T at 5 K. This indicates that noncollinear spins exist at the interfaces between two phases, creating a competition between interfacial Dzyaloshinskii-Moriya (DM) and exchange interactions. Strikingly, in addition to the normal Hall effect by Lorentz force and anomalous Hall effect by magnetization, we observe new contribution namely topological Hall effect below 75 K. This verifies the existence of topological spin texture, which is the consequence of competing interactions controlled by both applied field and temperature. Our work demonstrates that spintronic devices may be fabricated exploiting rich magnetic properties of different phases.

## I. Introduction

Most magnetic materials exhibit ferromagnetic (FM) or antiferromagnetic (AFM) collinear order as the consequence of exchange interaction. However, the broken inversion symmetry at interfaces of two different phases may give rise to an interfacial Dzyaloshinskii-Moriya (DM) interaction, which favors spin canting towards each other and leads to noncollinear spin texture.[1] An important phenomenon at the interface of FM and AFM materials is exchange bias (EB), which manifests itself in a shift of hysteresis loop along the magnetic field axis.[2,3] EB has been utilized to control the direction of magnetization in modern spintronics devices such as read heads in hard disk drives and memory elements in magnetic random access memory.[4,5] Most of theories which have been developed to unveil the details of EB assume uncompensated spins of an antiferromagnet at the interface to pin the ferromagnet. These theories therefore fail to explain the origin of EB in a system with a magnetically compensated interface.[3] Interfacial DM interaction has been proposed as a possible origin for EB in some compensated FM/AFM bilayers, such as $Fe_3O_4$/CoO [6] and $IrMn_3$/Co [7].

A particularly notable consequence of complex spin texture induced by the interfacial DM interaction is the interplay between charge and spin transport. In a noncollinear spin configuration the scalar spin chirality $\kappa = S_i \cdot (S_j \times S_k)$ (where $S_i$, $S_j$, and $S_k$ are local spins) can induce a finite Berry phase and an associated fictitious magnetic flux, giving rise to so-called topological Hall effect (THE).[8] In the past several years, there has been considerable activity in identifying and understanding THE seen in skyrmions [9,10], ferromagnetic/paramagnetic bilayer [11], topological insulator heterostructures [12], and non-coplanar antiferromagnets [13]. While THE is attributed to the special spin configurations, the essential ingredients to generate desired spin textures are yet to be identified. For example, DM interaction [1] is considered to play an important role, which requires anti-symmetric crystal structure, anisotropic exchange coupling, and strong spin-orbit coupling. However, it was also pointed out that THE can exist in systems with little spin-orbit interaction [8], which may imply that the DM interaction is unimportant. In this study, we choose a manganese nitride system with mixed ferrimagnetic (FI) ε-phase $Mn_4N$ and AFM ζ-phase $Mn_2N_y$, which is different from any previously reported heterostructures exhibiting THE. This results in the EB effect with a giant bias field $H_{EB}$ of ~ 0.22 T at 5 K. Moreover, we observe a noticeable THE in this FI/AFM system over a wide range of both temperature and magnetic field, which is the signature of the topological spin textures. We argue that THE is induced by the interfacial DM interaction between FI and AFM phases.

## II. Material Preparation & Characterization Methods

Manganese nitride films with the thickness of 150 nm were grown on MgO (001) substrates using an Omicron customized multiprobe plasma-assisted molecular beam epitaxy (MBE) system. Before deposition, the substrate was cleaned *ex-situ* using solvent, first with acetone and then with isopropyl-alcohol. Additional *in-situ* cleaning of the substrate was performed by annealing at 1000 °C with nitrogen plasma incident for 1 h. Manganese flux was provided by a custom designed effusion cell operated ~ 900 °C, whereas N flux was supplied by an RF nitrogen Plasma Source. The optimal growth condition for single crystal ε-phase $Mn_4N$ has been determined to be at the deposition temperature of 450 °C and $N_2$ pressure of $9.5\times10^{-6}$ mbar with the radio frequency power supplied to electron cyclotron resonance of 300 W.[14] After growth, the film was further annealed at 510 °C in vacuum for 1 hour to introduce the ζ-phase $Mn_2N_y$. The growth was monitored by *in-situ* reflection high-energy electron diffraction. The microstructure was investigated by X-ray diffraction (XRD) with Cu Kα radiation. The magnetic properties were measured by Magnetic Property Measurement System (MPMS - 7 T, *Quantum Design*). The electrical transport measurements were carried out in a Physical Properties Measurement System (PPMS - 14 T, *Quantum Design*).

## III. Results & Discussion

Manganese nitride ($Mn_xN_y$) forms a variety of phases including θ-phase (MnN), η-phase ($Mn_3N_2$), ε-phase ($Mn_4N$) (see Fig. 1(a)), and ζ-phase ($Mn_2N_y$) (see Fig. 1(b)).[15] To characterize our films, we have performed XRD measurement. Fig. 1(c) plots the XRD *θ-2θ* scan at room temperature for the film after annealing at 510 °C. As indicated in the figure, both the ε-phase $Mn_4N$ (002) peak and a set of ζ-phase peaks were detected. The ε-phase $Mn_4N$ has the fcc anti-perovskite structure as illustrated in Fig. 1(a). Nitrogen is located at the body center and two inequivalent manganese sites occupy the corner (Mn(I)) and face centered (Mn(II)) positions with magnetic moments of 3.85 $\mu_B$/f.u. and 0.9 $\mu_B$/f.u along the *c* axis (see arrows in Fig. 1(a)), respectively. The FI transition temperature $T_{FI}$ is ~ 738 K.[16] As shown in Fig. 1(b), the ζ-phase $Mn_2N_y$ is formed by nitrogen atoms inserted into the hexagonal closely packed (hcp) Mn metal lattice. However, the octahedral sites are only partially occupied by nitrogen atoms with 0.16 < y < 0.72.[17] The ζ-phase $Mn_2N_y$ orders antiferromagnetically with 1.7$\mu_B$ per Mn along the *c* axis (see arrows in Fig. 1(b)).[18] For our film, the XRD shows peaks corresponding to three different planes of the ζ-phase, indicating it is not epitaxial. For comparison, the XRD pattern for the as-grown film is shown in the inset of Fig. 1(c), revealing signal from the ε-phase only. This indicates that the ζ-phase is indeed induced by the post annealing.

Fig. 1(d) shows the temperature dependence of magnetization (M) measured by applying magnetic field $H$ = 0.1 T, normal to the plane of the film. With decreasing temperature, the magnetization in both zero-field-cooling (ZFC) and field-cooling (FC) modes increases slowly at high temperatures, but raises dramatically below 20 K. This is mainly duo to the MgO substrate, which exhibits very strong paramagnetism at low temperatures. This also makes it difficult to identify any magnetic phase transition in our film.

Fig. 1(e) displays the temperature dependence of the resistivity $\rho$ of the mixed-phase film. For comparison, the resistivity for the pure ε-phase film is also shown as a dash line. While the positive slope indicates that the film is metallic, the resistivity is more than one order higher than that of the ε-phase (dash line), indicating that the ζ-phase reduces the electric conduction. More importantly, d$\rho$/dT is very different for the processed film compared to the ε-phase, as shown in Fig. 1(f): it is larger for the processed film, and there is a clear signature of the AFM transition $T_N$ ~ 273 K for the ζ-phase $Mn_2N_y$, which is absent in the case of pure ε-phase. Below $T_N$, $\rho(T)$ varies more or less logarithmically, i.e., $\rho \propto \ln T$, as indicated in Fig. 1(f). The logarithmic temperature dependence of resistivity is usually observed in Kondo systems as the consequence of interaction between the itinerant electrons and local magnetic moment. While this scenario may occur at the interfaces in our film, the positive slope of ρ(T) is inconsistent with the Kondo picture. Nevertheless, ρ(T) shows a typical coherent Fermi liquid behavior below ~ 75 K. The black line in Fig. 1(f) is the fit of data using $\rho = \rho_0 + AT^2$, with $\rho_0$ = 132 μΩ cm and A = 0.006 μΩ cm/K$^2$. Note that there is a hump in dρ/dT around ~ 75 K, which is also seen in the pure ε-phase (see Fig. 1(f)). This indicates that zero-field electrical transport is mainly through the ε-phase in our film.

Magnetic property measurements clearly indicate the mixed magnetic phases of the film. Fig. 2(a) shows the magnetic hysteresis loops under three different conditions at 5 K, all with the external field applied perpendicular to the film. First, the film is demagnetized at 385 K from 5 T to zero field prior to zero-field cooling. This loop, defined as spontaneous, has a symmetric magnetic loop with respect to the origin. After cooling from 385 K under +2 Tesla, a prominent shift of the center of the loop along the magnetic field axis is observed towards negative fields. In contrast, on cooling in the presence of a -2 Tesla field, the loop is shifted in the positive direction. This behavior reveals the presence of exchange bias with a very large bias field $H_{EB}$ of ~ 0.22 T, defined as the absolute offset of the loops along the field axis. In addition, field-cooling loops have a much larger coercivity $H_C$ compared with that of the spontaneous loop. This enhancement of $H_C$ between field-cooling loops and zero field-cooling loop is the *signature of exchange bias* effect.

In Fig. 2(b), we show the characteristic magnetic hysteresis loops measured at indicated temperatures after field cooling from 385 K in the presence of a +2 Tesla field. Note that the hysteresis loop and corresponding saturation magnetization $M_S$ become smaller with increasing temperature. The inset shows that $M_S$ decreases with increasing T. Fig. 2(c) shows the temperature dependence of $H_{EB}$ and $H_C$ extracted from data shown in Fig. 2(b). The $H_{EB}$ monotonically decreases with increasing temperature, vanishing around 300 K. Typically, the blocking temperature $T_B$, i.e., the temperature of the vanishing exchange bias, is lower than the Néel temperature of the AFM component. In our case, $T_B$ is slightly higher than $T_N$ determined from resistivity (Figs. 1(e) and 1(f)), likely resulting from AFM fluctuation above $T_N$. Nevertheless, a symmetric magnetic loop is recovered above $T_B$ with reduced $H_C$ (see Fig. 2(c)). While $H_C$ increases with decreasing temperature in a wide temperature range, there is *clearly a dip centered at 35 K*, which will be discussed later.

The existence of exchange bias and non-monotonic behavior of $H_C(T)$ indicate that spins at FI/AFM interfaces play important role in magnetic properties. The dynamics of the spin structures at the interfaces between the two phases can be revealed by magnetic training. After cooling from 385 K under a +2 Tesla field, 14 sequencing magnetic loops were measured at 5 K. Compared with the first loop (inset of Fig. 2(d)), we find that the exchange bias field decreases as the loops are repeated, which is the characteristic of the effect of training. Fig. 2(d) is the measured exchange bias field as a function of hysteresis loop index n. $H_{EB}$ is found to decrease monotonically with increasing n, verifying spin rearrangement at the magnetically disordered interfaces.

For pure AFM spin rearrangement at the interface, the $H_{EB}(n)$ is proportional to $n^{-1/2}$.[19] However, this expression cannot describe our experimental data well. To explain the observed training effect, we adapt the model that assumes two different relaxation rates [20] with one from AFM and the other from rotatable spin components at the interface:

$$H_{EB}^n = H_{EB}^\infty + A_{AFM} \exp(-n/P_{AFM}) + A_R \exp(-n/P_R), \qquad (1)$$

where $A_{AFM}$ and $P_{AFM}$ are related to the change of the AFM spins, $A_R$ and $P_R$ are parameters denoted to the rotatable spins. This model fits the data as shown in Fig. 2(d) as a black dash line. The parameters obtained from the fit are $H_{EB}^n = 1137$ Oe, $A_{AFM} = 572$ Oe, $P_{AFM} = 0.26$, $A_R = 12473$ Oe, $P_R = 1.59$. The ratio $P_R/P_{AFM} \sim 6$ implies that the AFM spins rearrange (relax) nearly 6 times slower than the rotatable ones, confirming the importance of interfacial spin interaction.

To understand the interfacial magnetic properties, we have measured the Hall effect with a Hall bar configuration shown in Fig. 3(a). Fig. 3(b) displays the Hall resistivity $\rho_H$ as a function of magnetic field at indicated temperatures, each taken after cooling from 385 K under +2 Tesla. The rapid increase in $\rho_H$ in weak field indicates the rotation of domains into alignment with H. The hysteresis loop of $\rho_H$ is reminiscent of magnetization, indicating the existence of anomalous Hall effect (AHE). In this model, the Hall resistivity can be written as $\rho_H = R_o H + R_s M$, where $\rho_o$

= $R_oH$ is the ordinary term and $\rho_{AH} = R_SM$ is the M-linear anomalous term, where $R_o$ and $R_S$ are the ordinary and anomalous Hall coefficients, respectively. Above the saturation field $H_S$, $\rho_{AH}$ is constant and the small ordinary Hall component $R_oH$ is visible as a linear background. The anomalous Hall coefficient is related to the resistivity $\rho$, with $R_S = S_A\rho^2$, where constant $S_A$ is proportional to the spin-orbit interaction.[21] Since the change of $\rho$ under the transverse field (H $\perp$ I) is very small ( less than 1% at fields up to 5 T, see Fig. S1 in *Supporting Information*); $R_S$ can be treated as a constant under magnetic field. By linear fitting $\rho_H(H)$ in the high field region (see Fig. S2 in *Supporting Information*), $R_o$ and $\rho_{AH}$ ($S_A$) can be obtained from the intercept and the slope. Fig. 3(c) and Fig. 3(d) display $R_o$ and $\rho_{AH}$ as a function of temperature, respectively. Note that $R_o$ is positive above 75 K but becomes negative at lower temperatures. This suggests a change of electronic structure at ~ 75 K, below which the zero-field resistivity exhibits $T^2$ dependence (Fig. 1(f)). On the other hand, $\rho_{AH}$ increases with increasing temperature from 5 to 275 K and exhibits a maximum at 275 K, corresponding to $T_N$ of the ζ-phase $Mn_2N_y$ (Fig. 1(f)). Such temperature dependence of $\rho_{AH}(T)$ is different from that of $M_S$ shown in the inset of Fig. 2(b), indicating that the anomalous Hall effect is controlled by both FI and AFM phases.

Large deviation of $\rho_H$ from the magnetization can be found by comparing $\rho_H$ (red lines) and $R_oH+R_SM$ (blue lines) as shown in Fig. 4(a). At 5 K, in addition to the discrepancy in coercive field and exchange bias, there is another anomaly between ± 0.5 and ±2 T: a hump structure between 0.5 to 2 T in the ascending field curve but between -0.5 to -2 T in the descending field curve. Figs. 4(b)-4(c) show the evolution of this feature marked by shaded areas, which becomes visible below 75 K. Such a hump structure is seen in systems with *topological spin texture*.[11,12] In the presence of topological spin texture, the Hall resistivity $\rho_H$ is expressed as

$$\rho_H = R_oH + R_SM + \rho_{TH}, \qquad (2)$$

where $\rho_{TH}$ is the topological Hall resistivity. Following the procedure in literature [11,12], we quantify $\rho_{TH}$ using the hump structure area between the ascending and descending curves (indicated as the shaded area in Figs. 4(b)-4(c)) marked $\Delta\rho_H$. Fig. 4(d) shows the color map of $\Delta\rho_H$ in the H-T plane, which indicates (1) the topological spin state below 75 K and (2) the largest intensity of THE is located around 35 K and +1 T via ascending field, and -1.5 T via descending field. Recalling that both $H_C$ (Fig. 2(c)) and $M_S$ (inset of Fig. 2(d)) reach the local minimum at 35 K, we believe that the *THE is the consequence of competing interactions between AFM and FI phases at the interfaces, giving rise to optimal condition to form topological spin texture around 35 K*. The THE intensity difference between positive and negative fields should be caused by exchange bias. Remarkably, the range of temperature and field for the presence of THE in our film is much wider than that in the skyrmions [11,12,22-24], suggesting much larger driving force for spin rearrangement in our film.

Given that the hysteresis loop of $\rho_H(H)$ is different from that of M(H), the exchange bias from these two quantities is expected to be different. Fig. 4(e) shows $H_{EB}$ extracted from $\rho_H(H)$ and M(H) hysteresis loops, which deviate from each other below 75 K, when topological spin structure is formed. Note that we measured M(H) and $\rho_H(H)$ from the same sample which experienced the same magnetic field cooling procedure. In addition, we measured M(H) loops twice, before and after patterned the film into a Hall bar, which show the same behavior. Therefore, the observed difference in $H_{EB}$ is intrinsic, which reflects different response of the Hall effect and M to the topological spin structure. As mentioned previously, $R_o$ changes sign ~ 75 K, suggesting the modification of electronic structure in this multiband system. On the other hand, in the presence of the spin-orbit interaction $\lambda_{SO}$, the net total chirality of the rotatable spins with the noncollinear spin textures couples to the net magnetization $M$, [25] leading to the chirality-driven THE, whose magnitude should be proportional to $M$ and $\lambda_{SO}$ (where coupling coefficient depends on the detailed band structure).[26] The larger $H_{EB}$ observed in $\rho_H(H)$ may imply that, additional scattering, for example with AFM spins, plays an important role, in addition to scattering associated with rotatable spins, while exchange bias in M(H) more from rotatable spins than the AFM spins.

## IV. Conclusion

To summarize, we have successfully fabricated manganese nitride films consisting of both the ferrimagnetic phase (ε-phase) and the antiferromagnetic phase (ζ-phase). This allows us to observe novel phenomena attributable to the interfaces between FI and AFM phases: (1) finite exchange bias in the magnetization and Hall resistivity below $T_N$ of the AFM phase, and (2) topological Hall effect below 75 K with the largest signal centered around 35 K. The exchange bias extracted from the Magnetization reaches ~ 0.22 T at 5 K, much larger than other systems. The topological Hall effect reflects the formation of topological spin texture, which is the consequence of competing between interfacial DM and exchange interactions. These interactions depend on both applied field and temperature. Our work paves a way for utilizing their rich magnetic properties of different phases to fabricate desired spintronic devices.

**Acknowledgments**

Work at LSU is supported by the US National Science Foundation via grant DMR-1504226. MM acknowledges financial support from China Scholarship Council.

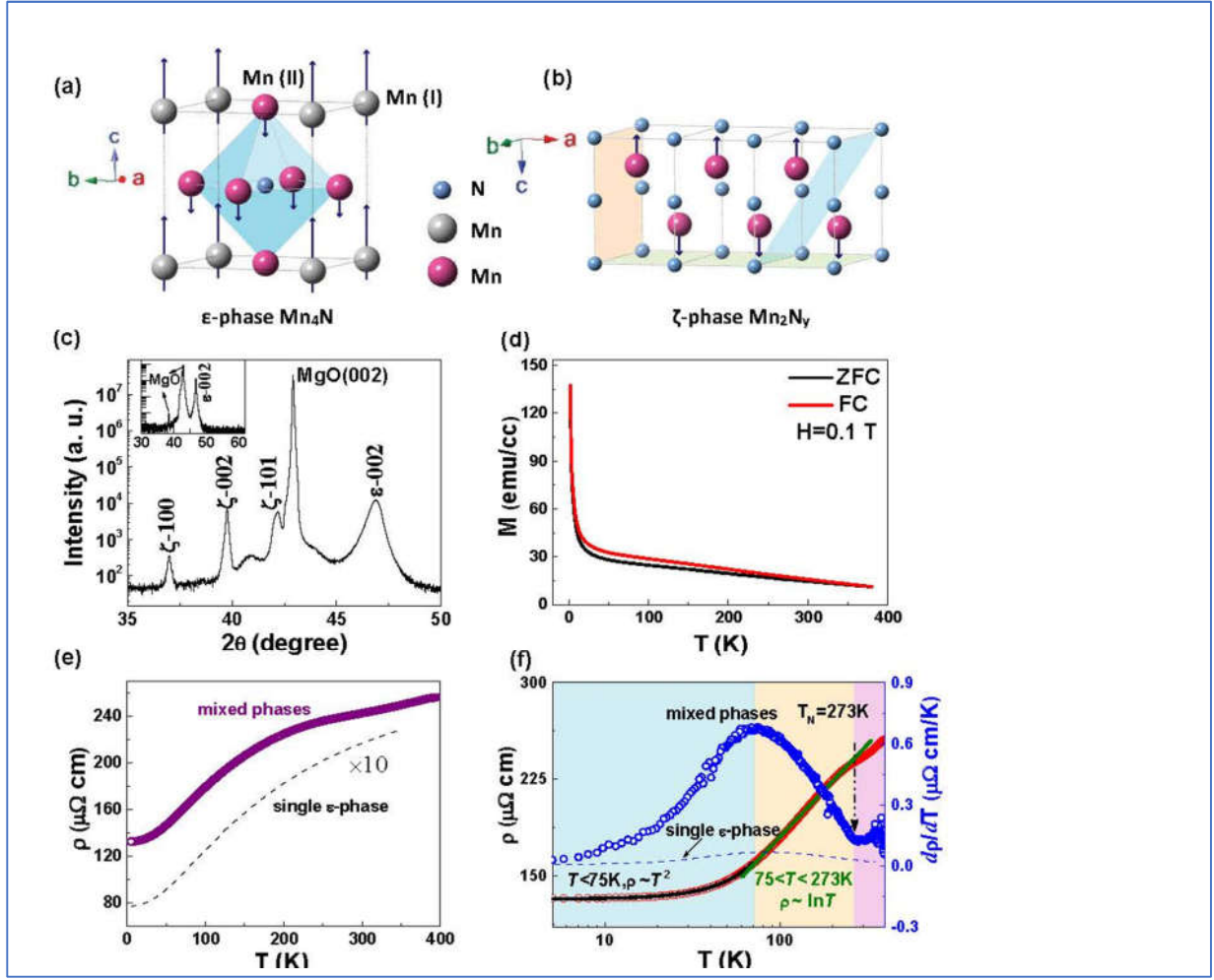

**Fig. 1.** (a) Schematic of antiperovskite $Mn_4N$ ε-phase structure and magnetic alignment (black arrows) below 738 K. (b) Sketch of ζ-phase $Mn_2N_y$ and magnetic structure (black arrows). Orange, blue, and green planes correspond to (100), (101), and (002) peaks observed in XRD, respectively. (c) X-ray diffraction of the $Mn_xN_y$ film after annealing at 510 ºC. Inset: X-ray diffraction of the as-grown ε-phase film. (d) Temperature dependence of magnetization under field cooling (FC) and zero field cooling (ZFC) measured by applying magnetic field H= 0.1 T, normal to the film plane. (e) T dependence of longitudinal resistivity ρ for mixed phases $Mn_xN_y$ (purple circles) and for single ε-phase film with 10 times magnification (dash line). (f) Semi-logarithmic plot of ρ(T) (red) and its derivative dρ/dT(T) (blue). For comparison, dρ/dT(T) for the ε-phase film is also plotted (dash line). The black solid line represents a fit using $\rho = \rho_0 + AT^2$ below 75 K and the green line shows that ρ has a lnT relation between 75 and 273 K.

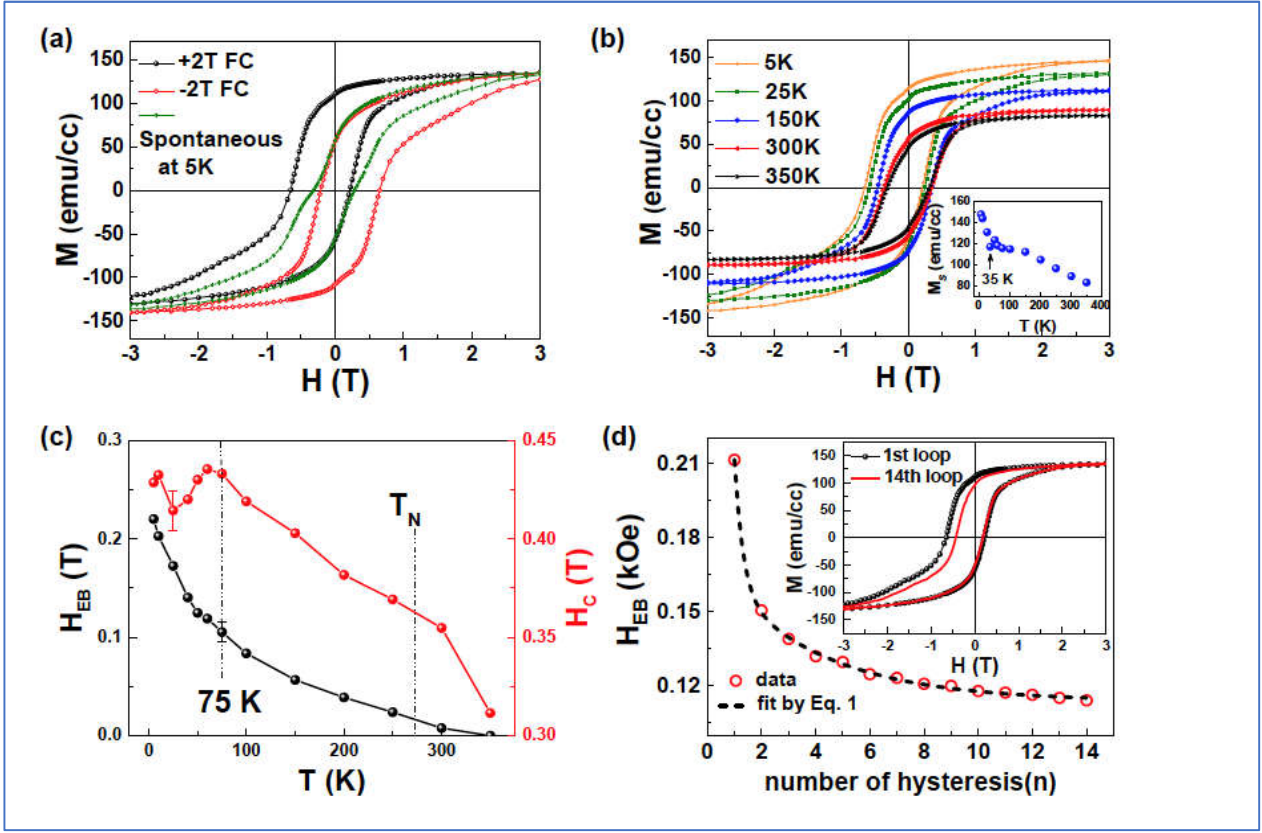

**Fig. 2.** (a) Magnetic hysteresis loops measured at 5 K after FC from 385 K in the presence of a +2 T (black solid spheres), -2 T (red open squares) magnetic field, and ZFC from 385 K (green open circles). (b) Magnetic hysteresis loops at indicated temperatures each taken after FC from 385 K in the presence of +2 T magnetic field. Inset shows saturation magnetization $M_S$ taken at H = 3 T as a function of T. (c) T dependence of exchange bias field ($H_{EB}$) and coercive field ($H_C$). (d) $H_{EB}$ as a function of hysteresis cycles (n) at 5 K. The dash line is the fit of data using Eq. (1). Inset shows the first and 14th loops.

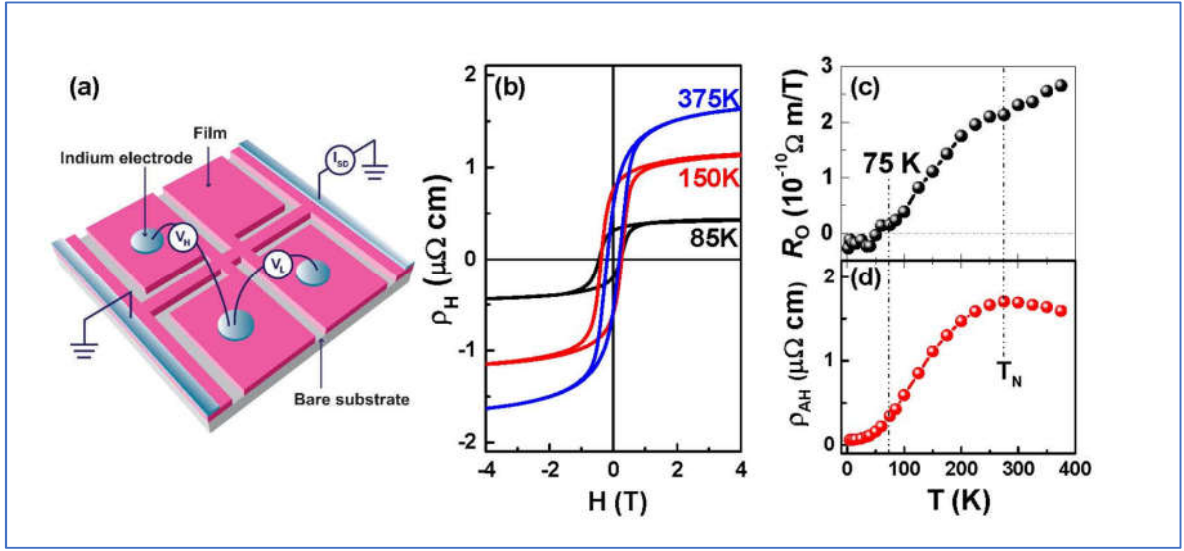

**Fig. 3**. (a) Schematic drawing of a sample for the Hall effect measurement. (b) Magnetic field dependence of Hall resistivity at indicated temperatures. Ordinary Hall coefficient $R_o$ (c) and anomalous Hall resistivity $\rho_{AH}$ (d) as a function of T.

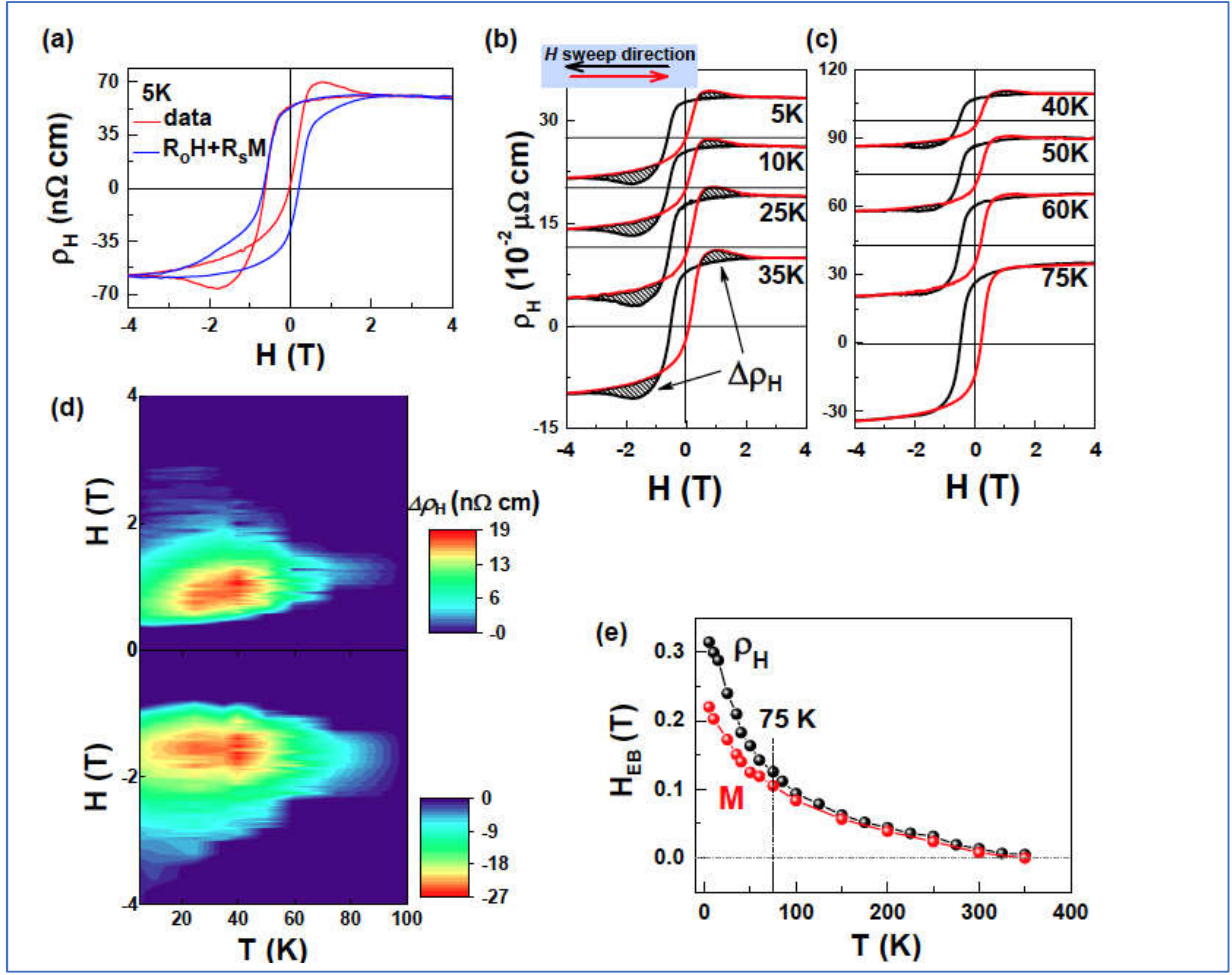

**Fig. 4**. (a) Comparison of $\rho_H$ with calculated $R_oH + R_SM$ at 5 K. (b - c) $\rho_H$ loops below 75 K. The difference between the ascending (red line) and descending (black line) field curves is shaded and marked as $\Delta\rho_H$. (d) Contour map of $\Delta\rho_H$ in the H - T plane. (e) T dependence of $H_{EB}$ deduced from $\rho_H(H)$ loops and M(H) loops.

# Supporting Information for " Observation of Giant Exchange Bias and Topological Hall Effect in a Manganese Nitride Film"

Meng Meng,[1,2] Shuwei Li,[1] Mohammad Saghayezhian,[2] E. W. Plummer,[2] Rongying Jin,[2]

1. State Key Laboratory of Optoelectronic Materials and Technologies, School of Materials Science & Engineering, Sun Yat-sen University, Guangzhou 510275, China
2. Department of Physics & Astronomy, Louisiana State University, Baton Rouge, Louisiana 70803, USA

**Perpendicular magnetoresistance**

Fig. S1 shows the transverse magnetoresistance (MR) of the $Mn_xN_y$ film at 5 K. Alignment of the ferrimagnetic ordering by magnetic field makes the scattering more coherent, thus the MR is negative. However, the MR is small, reaching -0.2% under 5 T at 5 K. Since it is even smaller at higher temperatures, we ignore the field dependence in analyzing anomalous Hall effect between 2 and 4 T.

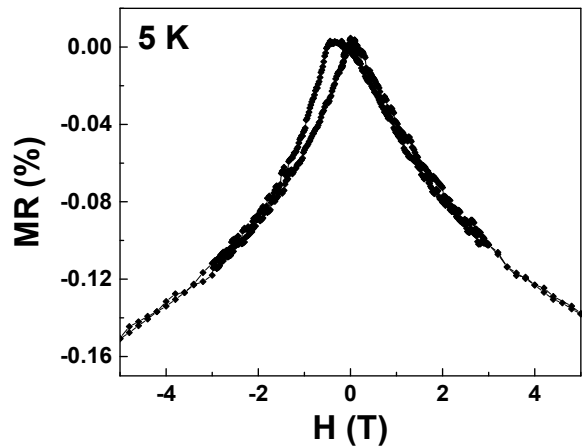

Fig. S1. Transverse magnetoresistance of the $Mn_xN_y$ film at 5 K.

**Determination of ordinary and anomalous Hall coefficients**

The Hall resistivity $\rho_H = R_o H + R_S M$ contains the ordinary term $R_o H$ and the anomalous term $\rho_{AH} = R_S M = S_A \rho^2 M$ where H is magnetic field perpendicular to the sample plane, $R_o$ and $R_S$ are the ordinary and anomalous hall coefficients, respectively. Above the saturation field $H_S$, $R_S M$ is constant (because the resistivity has little field dependence as shown in Fig. S1) and the ordinary Hall component $R_o H$ is visible as a linear background. Fig. S2(a) shows the field dependence of the Hall resistivity at 350 K. Above the saturation field $H_S$, the Hall resistivity is linear and its slope represents the ordinary Hall coefficient $R_o$. And the anomalous Hall resistivity $\rho_{AH}$ is determined as the y-intercept of a linear fit to the high-field regions. To obtain $R_o$ and $R_S$ directly, we replot the Hall data as $\frac{\rho_H}{H}$ versus $\frac{\rho^2 M}{H}$ in high field regions as shown in Fig. S2(b). In this plot, the intercept is $R_o$ and the slope is $S_A$.

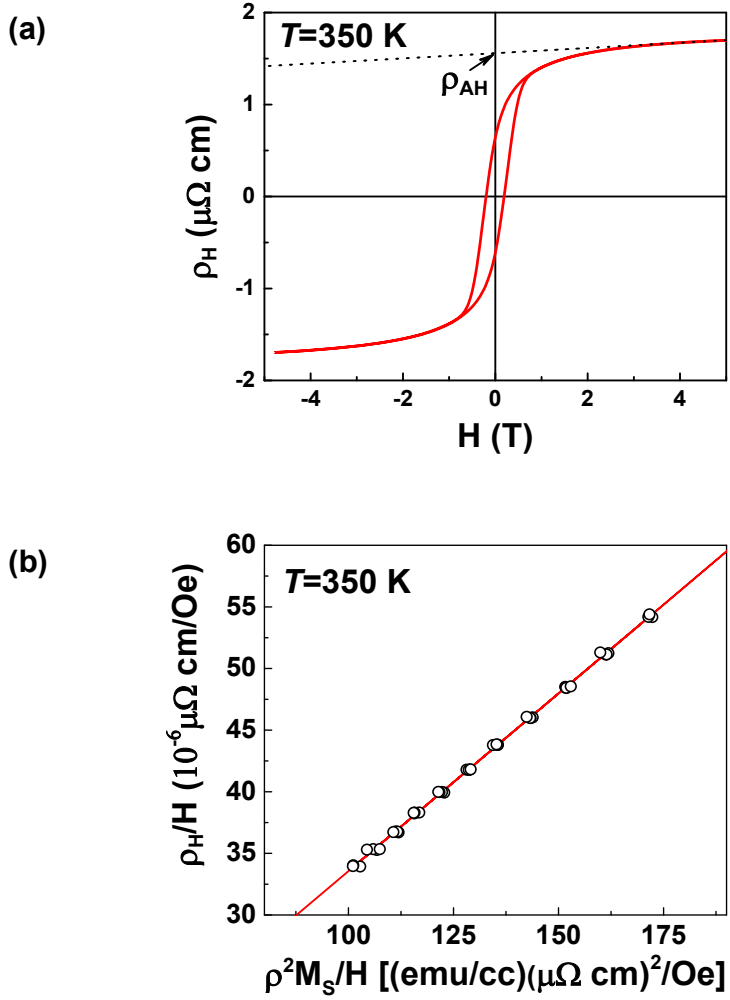

Fig. S2. (a) The total Hall resistivity $\rho_H$ (red line) and high field linear fit (dashed line) at 350 K. $R_o$ can be extracted from the slope of linear fit at high field. Anomalous Hall resistivity $\rho_{AH}$ is the y-intercept of the linear fit. (b) Plot of $\frac{\rho_H}{H}$ versus $\frac{\rho^2 M}{H}$ in high field regions yields a linear relationship. The slope and the y-intercept of the linear fit are coefficients $S_A$ and $R_o$, respectively. $R_o$ can also be determined from the method described in (a).